\documentstyle[aps,prb,multicol,revfig209]{revtex}
\begin{document}
\draft
\title{Interplay among critical temperature, hole-content,
and pressure in the cuprate superconductors}
\author{Giuseppe G.N. Angilella, Renato Pucci, and Fabio Siringo}
\address{Dipartimento di Fisica dell'Universit\`a di Catania,
57, Corso Italia, I-95129 Catania, Italy}
\date{Received 22 December 1995; revised manuscript received 17 July 1996}
\maketitle
\begin{abstract}
Within a BCS-type mean-field approach to the extended Hubbard 
model, a nontrivial dependence of $T_c$ on the hole content per 
unit CuO$_2$ is recovered, in good agreement with the celebrated 
non-monotonic universal behaviour at normal pressure. Evaluation 
of $T_c$ at higher pressures is then made possible by the 
introduction of an explicit dependence of the tight-binding band 
and of the carrier concentration on pressure $P$.
Comparison with the known experimental data for underdoped
Bi2212 allows to single out an `intrinsic' contribution to 
${\rm d} T_c /{\rm d} P$ from that
due to the carrier concentration, and provides a remarkable estimate 
of the dependence of the inter-site coupling strength on the lattice scale.
\end{abstract}
\begin{multicols}{2}
\section{Introduction}
The comparison between superconductive and normal state properties
of the high-temperature superconductors often
unveils quite remarkable features in their phenomenology, thus
helping in establishing most of the known universal behaviours.
Among them, the nontrivial dependence of the critical temperature
$T_c$ on the hole content $\rho$ per unit CuO$_2$
is probably the most celebrated.~\cite{Zhang1,Zhang2}

To that purpose, the high pressure data provide a natural
tool to investigate the correlations existing between the main
superconductive properties and the
structural properties, such as the relevant lattice
anisotropy, thus allowing to single out
the r\^ole of the carrier concentration.~\cite{Pucci,Klotz,Takahashi}

Unfortunately, earlier experimental works did not help in setting up
a homogeneous picture of the problem, possibly due to an uncareful
analysis of the data coming up from samples, often characterized by 
uneasily reproducible features, such as the hole content,
when even a slight presence of impurities 
has been proved to influence dramatically
the behaviour under pressure.~\cite{NeumeierC}

Nowadays there is a general conviction that pressure $P$ may affect $T_c$
both changing the hole content $\rho$, which is evidenced by
Hall resistance measurements,~\cite{Parker,Murayama}   
and in an `intrinsic' way, mainly due to 
a lattice rearrangement induced by pressure. 
Quantitatively, the latter statement can be summarized
by assuming $\rho =\rho(P)$ and $T_c = T_c (\rho, P)$, which 
yields~\cite{Schilling,Neumeier}
\begin{equation}
\label{eq:ansatz}
\frac{{\rm d} T_c}{{\rm d} P} = \frac{\partial T_c}{\partial P} +
\frac{\partial T_c}{\partial\rho} \frac{{\rm d}\rho}{{\rm d} P} .
\end{equation}
Since, for the majority of the compounds, Hall resistance
measurements under pressure 
suggest that ${\rm d}\rho/{\rm d} P >0$,~\cite{comment}
the different signs in ${\rm d} T_c /{\rm d} P$ 
observed in different compounds may be explained
as the result of a competition between the `intrinsic' contribution
$\partial T_c /\partial P$ and the known dependence of $T_c$
on $\rho$, through its pressure induced change, $\partial T_c /\partial \rho$,
which may be negative or positive for underdoped or overdoped samples
at normal pressure, respectively.
Different trends of $T_c$ as a function of $P$ have been
actually reported for several cuprate superconductors.~\cite{Takahashi} 
In particular, high pressure can improve $T_c$ up to an `optimal' value 
(${\rm d} T_c /{\rm d} P \geq 0$)
and then decrease it down to lower values 
(${\rm d} T_c /{\rm d} P < 0$).
By the way, possible extrinsic microscopic mechanisms have been recently
devised, in order to describe the pressure dependence of
the carrier concentration, especially due to oxygen ordering
effects, e.g., in Tl-based cuprates.~\cite{Klehe1,Klehe2,Klehe3}

The theoretical implications of such
a pressure depending $T_c$ and of such a wide range in the values of
${\rm d} T_c /{\rm d} P$ have been often 
considered mainly as checks to known 
theories,~\cite{Griessen,Wijngaarden,vanEenige}
though they are still inconclusive both on the nature of the
condensate pairs and on the nature and strength of their
coupling interaction.~\cite{Angilella}

In this paper, we shall address our attention to the generalization
at high pressures of a mean-field approach to a system of
interacting fermions. In Sec.~II, we shall outline 
the model and the mean-field approach at
normal pressure ($P=0$). The choice of a well-established tight-binding
dispersion relation for the carriers will make us able to reproduce
the observed dependence of $T_c$ on $\rho$, allowing a direct
comparison, e.g., 
with the experimental data available for Bi2212.~\cite{Allgeier}
In Sec.~III, we shall generalize such an approach to the case of an
applied external pressure ($P\neq 0$).
Reasonable pressure dependencies will be obtained
both for the band parameters and for the carrier concentration.
A comparison with available experimental data
for $T_c$ under pressures $P\leq 1.6$~GPa in underdoped
Bi2212~\cite{Huang} will be
presented, which will permit us to separate the two contributions to
${\rm d} T_c /{\rm d} P$ in Eq.~(\ref{eq:ansatz}). Besides,
a nontrivial dependence on pressure (and therefore on the lattice
scale) will be derived for the inter-site coupling strength,
thus suggesting a non-negligible lattice influence on the superconductive 
properties of the electronic system. We shall eventually 
summarize and address our conclusions in Sec.~IV.

\section{Mean-field approach at normal pressure}
\subsection{The model}
At normal pressure ($P=0$), in order to describe the interacting 
Fermi liquid of the hole-type carriers in an anisotropic lattice, we adopt an
extended Hubbard Hamiltonian
\begin{eqnarray}
H = && \sum_{\langle\langle ij\rangle\rangle\sigma} 
    t_{ij} c^{\dag}_{i\sigma} c_{j\sigma} 
  + U \sum_i n_{i\uparrow} n_{i\downarrow} \nonumber\\ 
  && + \sum_{\langle ij\rangle \sigma\sigma^\prime} V_{ij} 
    c^{\dag}_{i\sigma} c^{\dag}_{j\sigma^\prime}
    c_{j\sigma^\prime} c_{i\sigma} ,
\label{eq:HamiltonianR}
\end{eqnarray}
where $c^{\dag}_{i\sigma}$ ($c_{i\sigma}$) is a fermionic creation 
(annihilation) operator on the lattice site $i$, with spin projection
$\sigma\in\{\uparrow,\downarrow\}$ along a specified direction, and 
$n_{i\sigma} = c^{\dag}_{i\sigma} c_{i\sigma}$ is the density operator on site 
$i$. In Eq.~(\ref{eq:HamiltonianR}), $t_{ij}$ denotes the hopping 
integral between the lattice sites $\langle\langle ij\rangle\rangle$, located
at the positions $\bbox{R}_i$ and
$\bbox{R}_j = \bbox{R}_i + \bbox{\delta}_2$, 
respectively, with $\bbox{\delta}_2$ 
spanning over the vectors connecting a given site $i$ to its nearest-neighbour
and in-plane next-nearest-neighbour sites $j$
($t_{ij} \equiv t_{\bbox{\delta}_2}$, for translational invariance),
 $U$ measures the on-site interaction, while $V_{ij}$ describes the 
interaction between in-plane nearest-neighbour sites $\langle ij\rangle$, 
separated by the vectors $\bbox{\delta}_1$ 
($V_{ij} \equiv V_{\bbox{\delta}_1}$).

Using the standard transformation to the momentum representation
\begin{equation}
c^{\dag}_{\bbox{k}\sigma} = \frac{1}{\sqrt{N}} \sum_i e^{i\bbox{k}\cdot
\bbox{R}_i} c^{\dag}_{i\sigma} ,
\end{equation}
being $N$ the total number of lattice sites, the Hamiltonian 
Eq.~(\ref{eq:HamiltonianR}) takes the form
\begin{eqnarray}
H = && \sum_{\bbox{k}\sigma} \varepsilon_{\bbox{k}}
    c^{\dag}_{\bbox{k}\sigma} c_{\bbox{k}\sigma}
  + \frac{U}{N} \sum_{{\bbox{k}\bbox{p}\bbox{q}}\atop n}
    c^{\dag}_{\bbox{k}\uparrow} c^{\dag}_{\bbox{p}\downarrow}
    c_{\bbox{p}-\bbox{q}\uparrow} 
    c_{\bbox{k}+\bbox{q}+\bbox{G}_n \downarrow} \nonumber\\
  &&+ \frac{1}{N} 
    \sum_{{\bbox{k}\bbox{p}\bbox{q}}\atop{\sigma\sigma^\prime n}}
    \tilde{V}_{\bbox{q}}
    c^{\dag}_{\bbox{k}\sigma} c^{\dag}_{\bbox{p}\sigma^\prime}
    c_{\bbox{p}-\bbox{q}\sigma^\prime} 
    c_{\bbox{k}+\bbox{q}+\bbox{G}_n \sigma} ,
\label{eq:HamiltonianK}
\end{eqnarray}
where
\begin{equation}
\varepsilon_{\bbox{k}} = \sum_{\bbox{\delta}_2} 
t_{\bbox{\delta}_2} 
e^{i\bbox{k}\cdot{\bbox{\delta}_2}}
\label{eq:dispersionK}
\end{equation}
is the dispersion relation for the free carriers, in the tight-binding 
approximation, and
\begin{equation}
\tilde{V}_{\bbox{q}} = \sum_{\bbox{\delta}_1} 
V_{\bbox{\delta}_1} 
e^{i\bbox{q}\cdot{\bbox{\delta}_1}}
\label{eq:potentialK}
\end{equation}
is the Fourier transform of the nearest-neighbours inter-site interaction 
potential $V_{ij}$. In Eq.~(\ref{eq:HamiltonianK}), the sums over momenta 
run over the first Brillouin zone 
[e.g., $-\pi \leq k_i a_i <\pi$, $i=x,y,z$, for the
momentum $\bbox{k}$, being $a_i$ the spacings of an orthorhombic 
(nearly tetragonal) lattice
(Tab.~\ref{tab:Bi2212} for Bi2212)~\cite{Massidda1+Yu1}], and
momentum conservation is enforced up to a vector $\bbox{G}_n$ of the 
reciprocal lattice.

A few comments are now in order about the free dispersion relation,
Eq.~(\ref{eq:dispersionK}), and the interaction terms in the Hamiltonian,
Eq.~(\ref{eq:HamiltonianK}).

Detailed band structure calculations
indicate that the layered pattern of the cuprate oxides is
reflected in the charge density surrounding their lattice
sites.~\cite{Massidda1+Yu1,Yu2,Massidda2+Yu3} Such density exhibits a 
quasi-bidimensional arrangement, which closely follows the rich orbital
structure of the Cu and O ions.~\cite{Lopez} This behaviour
suggests a strong degree of hybridization along the bond
directions. The intermediate oxygens therefore provide suitable
bridgings between two nearest-neighbour coppers, thus favouring
directional charge transport. 
A quite complex band structure results, almost dispersionless in the
symmetry direction orthogonal to the Cu-O planes 
(Fig.~\ref{fig:dispersion}). Such a band structure gives rise to
exotic Fermi surfaces, which exhibit
quasicilindrical shapes, at typical fillings.~\cite{Siringo}

A tight-binding approximation to the band dispersion relation can be
employed up to an arbitrary degree of accuracy, by including a suitable
number of 
$\bbox{k}$-harmonics.~\cite{Annett,Norman}
Equation~(\ref{eq:dispersionK}) restricts to the lowest orders, and yields the 
model dispersion 
relation~\cite{Siringo,Schneider1+DeRaedt,Schneider2,Schneider3}
\begin{eqnarray}
\varepsilon_{\bbox{k}} &=& -2 t_x \cos (k_x a_x )- 2 t_y \cos (k_y a_y ) 
\nonumber\\
&& + 4 t_{xy} \cos (k_x a_x )\cos (k_y a_y ) -2 t_z \cos (k_z a_z ) -\mu ,
\label{eq:dispersion}
\end{eqnarray}
where $\mu$ denotes the Fermi level. The
hopping parameters $t_x$, $t_y$, $t_z$, $t_{xy}$ have been evaluated 
by comparison with the available angle-resolved photoemission 
spectra~\cite{Dessau,Ma} (ARPES) for the observed 
dispersion and Fermi surface in 
Bi2212~\cite{Massidda1+Yu1,Norman,Schneider2,Schneider3}
(Tab.~\ref{tab:Bi2212}). In particular, providing $t_z$ with a nonzero,
though small ($t_z \ll t_x ,t_y$) value, i.e., assuming a true, 
three-dimensional dispersion function
$\varepsilon_{\bbox{k}}$, ensures against the awkward
occurrence of van~Hove 
singularities. Figure~\ref{fig:dispersion} displays
Eq.~(\ref{eq:dispersion}) along a 
symmetry contour of the first Brillouin zone for
increasing pressure $P$ (cfr. {\em infra}). 

Since only states next to the Fermi surface do significantly contribute to the 
sums in the interaction terms of Eq.~(\ref{eq:HamiltonianK}), only the terms 
with $\bbox{G}_n =0$, $\bbox{k} + \bbox{p} = 0$ can be safely retained.
This eventually simplifies the Hamiltonian,
Eq.~(\ref{eq:HamiltonianK}), as
\begin{equation}
H = \sum_{\bbox{k}\sigma} \varepsilon_{\bbox{k}}
    c^{\dag}_{\bbox{k}\sigma} c_{\bbox{k}\sigma}
  + \frac{1}{N} \sum_{\bbox{k}\bbox{k^\prime}} V_{\bbox{k}\bbox{k^\prime}}
    c^{\dag}_{\bbox{k}\uparrow} c^{\dag}_{-\bbox{k}\downarrow}
    c_{-\bbox{k^\prime}\downarrow} 
    c_{\bbox{k^\prime}\uparrow} ,
\label{eq:Hamiltonian}
\end{equation}
where $V_{\bbox{kk^\prime}}$ includes the on-site interaction term and the 
restriction to the singlet channel only of the inter-site 
interaction~\cite{Leggett}
\begin{equation}
V_{\bbox{kk^\prime}} = U + \case{1}{2} 
(\tilde{V}_{\bbox{k}-\bbox{k^\prime}}
+ \tilde{V}_{\bbox{k}+\bbox{k^\prime}} ),
\end{equation}
which takes on the `separable' form
\begin{eqnarray}
V_{\bbox{kk^\prime}} = 
   U
   &&+ 2 V \cos (k_x a_x ) \cos (k_x^\prime a_x ) \nonumber\\
   &&+ 2 V \cos (k_y a_y ) \cos (k_y^\prime a_y ),
\label{eq:potential}
\end{eqnarray}
where a symmetric inter-site coupling constant $V$ has been assumed along both
directions in the Cu-O planes.

Such an interaction does not refer to any particular pairing mechanism,
and therefore does not require any particular nature of the couples.
However, experimental indications on the momentum dependence of the gap
function, although still questioned, clearly suggest an intermediate range
for the effective (renormalized) interaction between the carriers in the cuprate
planes. In particular, at least an inter-site attractive term is required, in order to allow
for a gap which displays the observed nodes in the ${\bbox k}$-space. Besides, the competition
between an attractive effective inter-site $V$ and a repulsive effective on-site
$U$ is expected to rule on the actual onset of superconductivity and the opening
of a gap. For our purposes, it is safe to retain only the on-site and inter-site
terms in the interaction, although other phenomenological properties of the cuprates
may suggest different functional forms for $\tilde{V}_{\bbox q}$.~\cite{Carbotte}

\btab
\begin{center}
\begin{tabular}{cccc}
$a_x$ & $a_y$ & $a_z$ & [\AA] \\ \hline
$5.414$ & $5.418$ & $30.89$ & \\
\end{tabular}
\begin{tabular}{ccccc}
$\kappa_x$ & $\kappa_y$ & $\kappa_z$ & $\kappa_V$ & [$10^{-3}$~GPa$^{-1}$ 
]\\ \hline
$4.3$ & $4.3$ & $8.3$ & $16.6$ & \\
\end{tabular}
\begin{tabular}{ccccc}
$t_x$ & $t_y$ & $t_z$ & $t_{xy}$ & [$e$V] \\ \hline
$0.05$ & $0.05$ & $0.005$ & $0.0225$ \\
\end{tabular}
\end{center}
\caption{Lattice parameters,~\protect\cite{Massidda1+Yu1} isothermal
compressibilities~\protect\cite{Cornelius} and 
band parameters~\protect\cite{Schneider2,Schneider3}
for Bi2212 at normal pressure ($P=0$).
}
\label{tab:Bi2212}
\etab

\bfig
\begin{center}
\includifigura{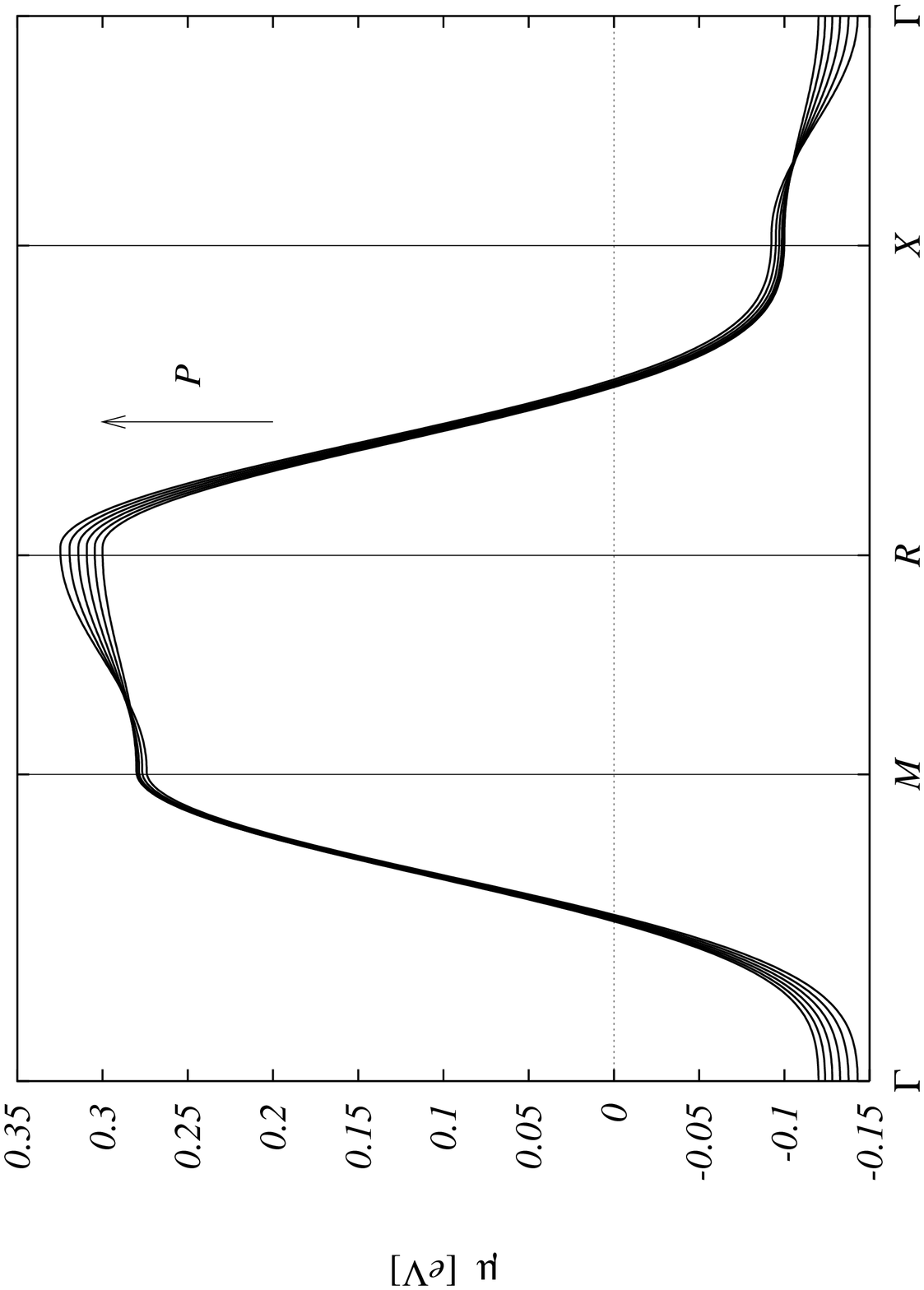}%
\end{center}
\caption{Band dispersion, Eq.~(\protect\ref{eq:dispersion}), along a
symmetry contour of the first Brillouin zone of a simple cubic lattice.
The different solid lines refer to increasing pressure,
$P=0\div 20$~GPa.
Due to the lattice anisotropic structure,
a weaker dispersion is recognized along
the directions $M$--$R$, $X$--$\Gamma$, parallel to the $z$ axis,
which remarkably increases by increasing pressure.}
\label{fig:dispersion}
\efig

\subsection{The approach}

In order to study the possibility for the Hamiltonian, 
Eq.~(\ref{eq:Hamiltonian}),
to give rise to a superconducting instability, a BCS-type mean-field
approximation can be employed. The superconducting condensate is then
characterized by an auxiliary gap field, which at finite
temperature $T$ obeys the BCS-type
self-consistent, nonlinear equation
\begin{equation}
\label{eq:gap}
\Delta_{\bbox{k}} = - \frac{1}{N} \sum_{\bbox{k^\prime}} 
V_{\bbox{kk^\prime}} F_{\bbox{k^\prime}} ,
\end{equation}
where
\begin{equation}
F_{\bbox{k}} = \frac{\Delta_{\bbox{k}}}{2E_{\bbox{k}}} 
\tanh\frac{E_{\bbox{k}}}{2T} ,~~~
E_{\bbox{k}} = \sqrt{\varepsilon^2_{\bbox{k}} + \Delta^2_{\bbox{k}}},
\end{equation}
and where we set Boltzmann's constant $k_{\rm B} =1$.
At the same degree of approximation,
the density of carriers $\rho$ is correspondingly given by~\cite{Leggett}
\begin{equation}
\label{eq:rho}
\rho(\mu,T) = \frac{1}{2N} \sum_{\bbox{k}} \left(
1 - \frac{\varepsilon_{\bbox{k}}}{E_{\bbox{k}}} 
\tanh \frac{E_{\bbox{k}}}{2T} \right),
\end{equation}
with $0\leq \rho\leq 1$. 
Although a mean-field approach could be generally questioned, since it does not 
take into account the `correlations,' nonetheless it proved itself a very 
useful approximation even in the case of non separable potentials.~\cite{Carbotte} 
Besides, standard diagrammatic techniques (though perturbative in nature) have
been recently employed in order to evaluate the corrections to the gap
due to the correlations in a simplified version of the model here 
employed.~\cite{Kuchiev} Such corrections, however, show up to be negligible 
in the strong coupling limit.

Due to the separable form of the potential, Eq.~(\ref{eq:potential}), 
direct inspection yields
\begin{equation}
\Delta_{\bbox{k}} = \Delta_0 + 2\Delta_x \cos (k_x a_x )+ 2\Delta_y \cos (k_y a_y ),
\end{equation}
being $\Delta_0$, $\Delta_x$, and $\Delta_y$ real constants satisfying the following
set of nonlinear, coupled equations
\begin{mathletters}
\label{eq:gapc}
\begin{eqnarray}
\Delta_0 + U \langle \Delta_{\bbox{k}}                 \rangle &=& 0,\\
\Delta_x + V \langle \Delta_{\bbox{k}} \cos (k_x a_x ) \rangle &=& 0,\\
\Delta_y + V \langle \Delta_{\bbox{k}} \cos (k_y a_y ) \rangle &=& 0,
\end{eqnarray}
\end{mathletters}
where
\begin{equation}
\langle f_{\bbox{k}} \rangle = \frac{1}{N} \sum_{\bbox{k}} \frac{f_{\bbox{k}}}{2E_{\bbox{k}}}
\tanh \frac{E_{\bbox{k}}}{2T} .
\end{equation}
The invariance property $\Delta_{\bbox{k}} = \Delta_{-\bbox{k}}$ follows again 
from the restriction of the inter-site interaction Eq.~(\ref{eq:potential}) to 
the singlet channel only.

Setting
\begin{equation}
\Delta_\pm = \case{1}{2} (\Delta_x \pm \Delta_y ) , 
\end{equation}
and forming the linear combinations
\begin{mathletters}
\begin{eqnarray}
S_{\bbox{k}} &&= \cos(k_x a_x ) + \cos(k_y a_y ) \\
D_{\bbox{k}} &&= \cos(k_x a_x ) - \cos(k_y a_y )
\end{eqnarray}
\end{mathletters}
the gap function can be re-expressed as
\begin{equation}
\Delta_{\bbox{k}} = \Delta_0 + 2\Delta_+ S_{\bbox{k}} + 2\Delta_- D_{\bbox{k}} .
\end{equation}

One may observe that the two sets of $\bbox{k}$-functions $\{ 1,S_{\bbox{k}} \}$
and $\{ D_{\bbox{k}} \}$ belong to two irreducible representations of the group
of rotations in the $(k_x , k_y )$ plane. In other words, one has
$\Delta_{\bbox{k}}
= \Delta_{\bbox{k}}^s + \Delta_{\bbox{k}}^d$, where $\Delta_{\bbox{k}}^s =
\Delta_0 + 2\Delta_+ S_{\bbox{k}}$ and $\Delta_{\bbox{k}}^d = 2\Delta_- D_{\bbox{k}}$,
which
explicitly display $s$-wave and $d$-wave symmetry character, respectively. Besides, all the dependence of
$\Delta_{\bbox{k}}$ on temperature $T$ and on the chemical potential $\mu$ is given by the set of the three
parameters $\Delta_0$, $\Delta_\pm$, whose relative value fix the overall symmetry pattern.

At zero temperature, and in the limiting case of $U=0$ ($\Delta_0 =0$),
Spathis et al.~\cite{Spathis} found 
that Equations~(\ref{eq:gapc}) can account for $s$ ($\Delta_x =\Delta_y$), 
$d$ ($\Delta_x = -\Delta_y$), as well as mixed $s$-$d$ ($|\Delta_x |
\neq |\Delta_y |$)
gap symmetry, depending on the position of the Fermi level
within the band. However, at the critical point ($T=T_c$), no symmetry
mixing is allowed,~\cite{Siringo} and two cases are possible: (i) $\Delta_- =0$,
and $\Delta_0 , \Delta_+ \to 0$ as $T\to T_c$ ($s$-wave);
(ii) $\Delta_0 , \Delta_+ =0$, and $\Delta_- \to 0$ as $T\to T_c$ ($d$-wave).
Therefore, if one lowers $T$ at a fixed chemical potential, one expects to observe first
a transition towards a superconducting state, characterized by a gap function of definite
symmetry, which can eventually evolve towards a mixed symmetry state, as $T$ lowers down to zero.

In either case, at $T=T_c$, it is possible to linearize Eqs.~(\ref{eq:gapc}) with respect to
$\Delta_0$, $\Delta_\pm$. A condition for the existence of a nontrivial solution is then easily
found to be
\begin{equation}
\label{eq:discr}
\biglb( 1+V\langle D_{\bbox{k}}^2 \rangle_c \bigrb)
\left[ \biglb( 1+U \langle 1\rangle_c \bigrb)\biglb(
1+U\langle S_{\bbox{k}}^2 \rangle_c \bigrb)
-UV \langle S_{\bbox{k}} \rangle^2_c \right] =0,
\end{equation}
where
\begin{equation}
\langle f_{\bbox{k}} \rangle_c = 
\lim_{T\to T_c} \langle f_{\bbox{k}} \rangle =
\frac{1}{N} \sum_{\bbox{k}} \frac{f_{\bbox{k}}}{2\varepsilon_{\bbox{k}}}
\tanh \frac{\varepsilon_{\bbox{k}}}{2T_c} .
\end{equation}

For fixed values of the coupling constants $U$, $V$ and of the chemical potential $\mu$,
Equation~(\ref{eq:discr}) yields, in general, two solutions for $T_c$. The larger one is
easily interpreted as the temperature below which a superconductive gap opens (with definite
symmetry), and the other as the temperature below which the gap symmetry mixing occurs. Moreover,
one of the two temperatures is clearly not affected by the presence of a non-zero on-site coupling,
$U$.

Figure~\ref{fig:Tc} displays $T_c$ vs $\rho$, consistently obtained through 
Eq.~(\ref{eq:rho}),
corresponding to the values $V\approx -0.052$~$e$V and $U=0.0 \div -V$ of the coupling parameters, obtained
by comparison with the available experimental data for Bi2212.~\cite{Allgeier}

As it can be seen, an $s$-wave gap opens at very low values of the hole-content $\rho$, corresponding
to low critical temperatures, whereas a $d$-wave gap is preferred near the optimal doping and beyond.
As expected, the on-site repulsion acts against the inter-site attraction with respect to
the onset of superconductivity: by increasing $U$ at constant $V$, one observes a decrease of 
$T_c$ (Fig.~\ref{fig:Tc}),
although the influence of $U$ is restricted only to the solution of 
Eq.~(\ref{eq:discr}) corresponding to an
$s$-wave gap. As a consequence of symmetry, the on-site interaction $U$ does not affect the
$d$-wave solution for the gap (see also 
Ref.~[\negthinspace\negthinspace\onlinecite{Siringo}] for a
full discussion). However, as Fig.~\ref{fig:Tc} clearly shows,
the comparison with the experimental results obtained by Allgeier and Schilling
in Ref.~[\negthinspace\negthinspace\onlinecite{Allgeier}] is reasonable only for the solution 
of Eq.~(\ref{eq:discr}), which corresponds to the opening of a $d$-wave gap. We can therefore safely restrict
ourselves to that case in the following.

The nontrivial
dependence of $T_c$ on the hole-like carrier concentration is
clearly non-monotonic, and correctly reproduces the qualitative
universal behaviour experimentally observed in the 
cuprates.~\cite{Zhang1,Zhang2}

\bfig
\begin{center}
\includifigura{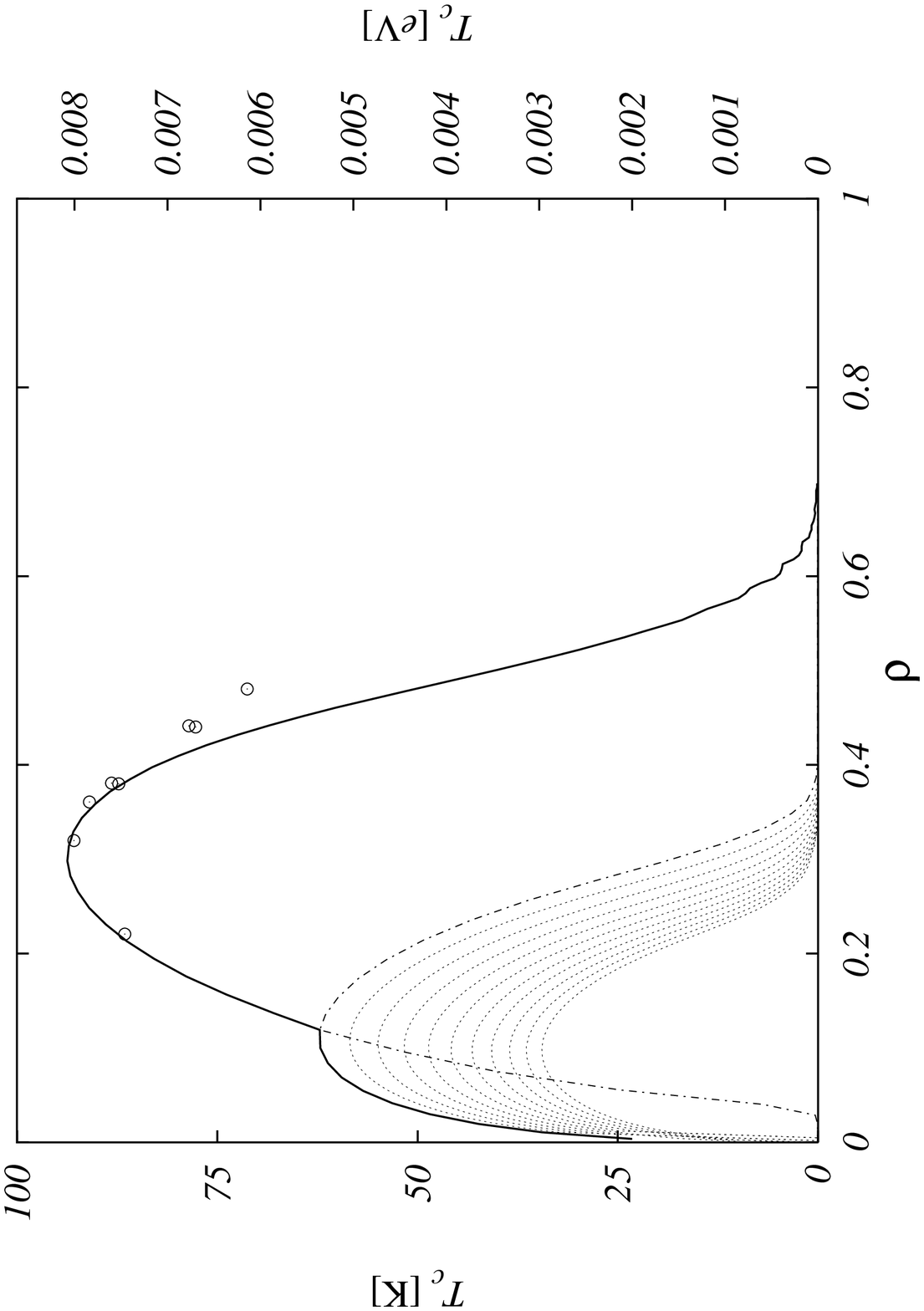}%
\end{center}
\caption{Critical temperature $T_c$ (solid line) and crossover
temperature towards gap symmetry mixing (dashed-dotted line)
vs hole-content $\rho$ at $P=0$. The effect of increasing
$U=0.0\div -V$ is also shown on the $T_c$ corresponding
to $s$-wave gap (dotted lines). By increasing $U$, one observes
a decrease of $T_c$ at fixed $\rho$. The circles are experimental
data obtained by Allgeier and Schilling~\protect\cite{Allgeier}
for Bi2212.}
\label{fig:Tc}
\efig

\section{Generalization to nonzero pressure}

The mean-field approach thus far described can be straightforwardly
generalized when a nonzero pressure $P$ is applied. The
effect of a pressure increase is threefold, involving both the
lattice and the carriers, either directly or indirectly: 
(i) it decreases the lattice
spacings and may distort the lattice itself, resulting in structural
phase transitions; 
(ii) it increases the carrier concentration, $\rho$; 
(iii) in oxygen doped cuprates, it may induce oxygen ordering, 
through a rearrangement of the excess ions into and from the Cu-O 
planes.~\cite{Klehe1,Klehe2,Klehe3}

At a given pressure $P$, the following nonzero (positive) components of
the isothermal compressibility tensor and 
of the isothermal volume compressibility may be defined:
\begin{equation}
\label{eq:compressibilities}
\kappa_i = -\frac{1}{a_i} \left( \frac{\partial a_i}{\partial P} \right)_T ,
~~~\kappa_V = - \frac{1}{V} \left( \frac{\partial V}{\partial P} \right)_T ,
\end{equation}
($i=x,y,z$), being $V=a_x a_y a_z$ the volume of an elementary cell.
Experimental values for $\kappa_i$ and $\kappa_V$
are listed in Tab.~\ref{tab:Bi2212} for Bi2212.~\cite{Cornelius}
Since they are almost constant over a quite wide range of pressure,
and since $\kappa_x \simeq \kappa_y$ (within the experimental error),
we may neglect transitions from the tetragonal to the orthorhombic structure.
Under such assumptions, 
at the lowest order in $P$
the lattice parameters are seen to obey a linear law:
\begin{equation}
\label{eq:aP}
a_i (P) = a_i (0) [1-\kappa_i P],
\end{equation}
which will be later used to parametrize the dependence of the lattice spacings on
pressure $P$.

Since $\kappa_i \approx 10^{-3}$~GPa (Tab.~\ref{tab:Bi2212}),
the lattice spacings keep their magnitude of several \aa{}ngstr\"oms, typical
of the cuprates, even at high pressures, so that
one may neglect the contribution to the band structure given by the ionic
core orbitals also at high pressures. We may therefore keep unchanged
the $\bbox{k}$-dependence of the tight-binding approximation to the
dispersion relation, Eq.~(\ref{eq:dispersion}), provided that
a pressure dependence is attributed to 
the hopping parameters $t_i$, $t_{xy}$.
Charge density calculations at normal pressure
suggest a very simple LCAO picture,~\cite{Massidda1+Yu1,Yu2,Massidda2+Yu3}
in which the hopping parameters are simply proportional
to the overlap integrals between
the main atomic orbitals involved in the
formation of the bonds between copper and oxygen sites.
The latter integrals can be easily worked out analytically as functions
of the lattice separations (see the Appendix for details).
Normalizing their values to those established at normal pressure,
Tab.~\ref{tab:Bi2212},
and making use of Eq.~(\ref{eq:aP}), we eventually obtain
a pressure dependent band dispersion relation.
Figure~\ref{fig:dispersion} displays the dispersion relation,
$\varepsilon_{\bbox{k}}$, Eq.~(\ref{eq:dispersion}),
along a symmetry contour of the first Brillouin zone, showing an
increasing dispersive behaviour in the direction orthogonal
to the Cu-O planes as $P$ increases. 
Figure~\ref{fig:dens} shows the densities of states (DOS)
\begin{equation}
n (\mu) = \sum_{\bbox{k}} \delta (\varepsilon_{\bbox{k}} -\mu),
\end{equation}
computed correspondingly as functions of the Fermi level.
As it can be seen, within this simple model, an applied pressure
widens the band extension, decreasing its bottom, $\mu_\bot$,
and increasing its top, $\mu_\top$,
and therefore lowers the height of what would have been a true van~Hove
singularity, so to keep $n$ normalized to unity,
\begin{equation}
\int_{\mu_\bot}^{\mu_\top} n(\mu){\rm d}\mu =1.
\end{equation}
The presence of such a large peak is mainly due to the quasi-bidimensional
character of the perovskite compounds, and it is confirmed both by
ARPES measurements~\cite{Dessau,Ma}
and by band structure calculations.~\cite{Schneider2,Schneider3}
Its relevance with respect to the phenomenological properties of the
high-$T_c$ superconductors has been recently underlined.~\cite{Plakida}
In particular, an antiferromagnetic-van~Hove (AFvH) theoretical 
picture~\cite{Dagotto1,Dagotto2} 
suggests a link between the presence of a large
peak in the DOS and the existence of an optimal value of $T_c$
at a small hole content.~\cite{Dagotto3} 
Since pressure decreases the height of the DOS
peak, that optimal hole content is expected to increase, which is
actually what we observed plotting $T_c$ vs $\rho$,
for increasing values of pressure.

\bfig
\begin{center}
\includifigura{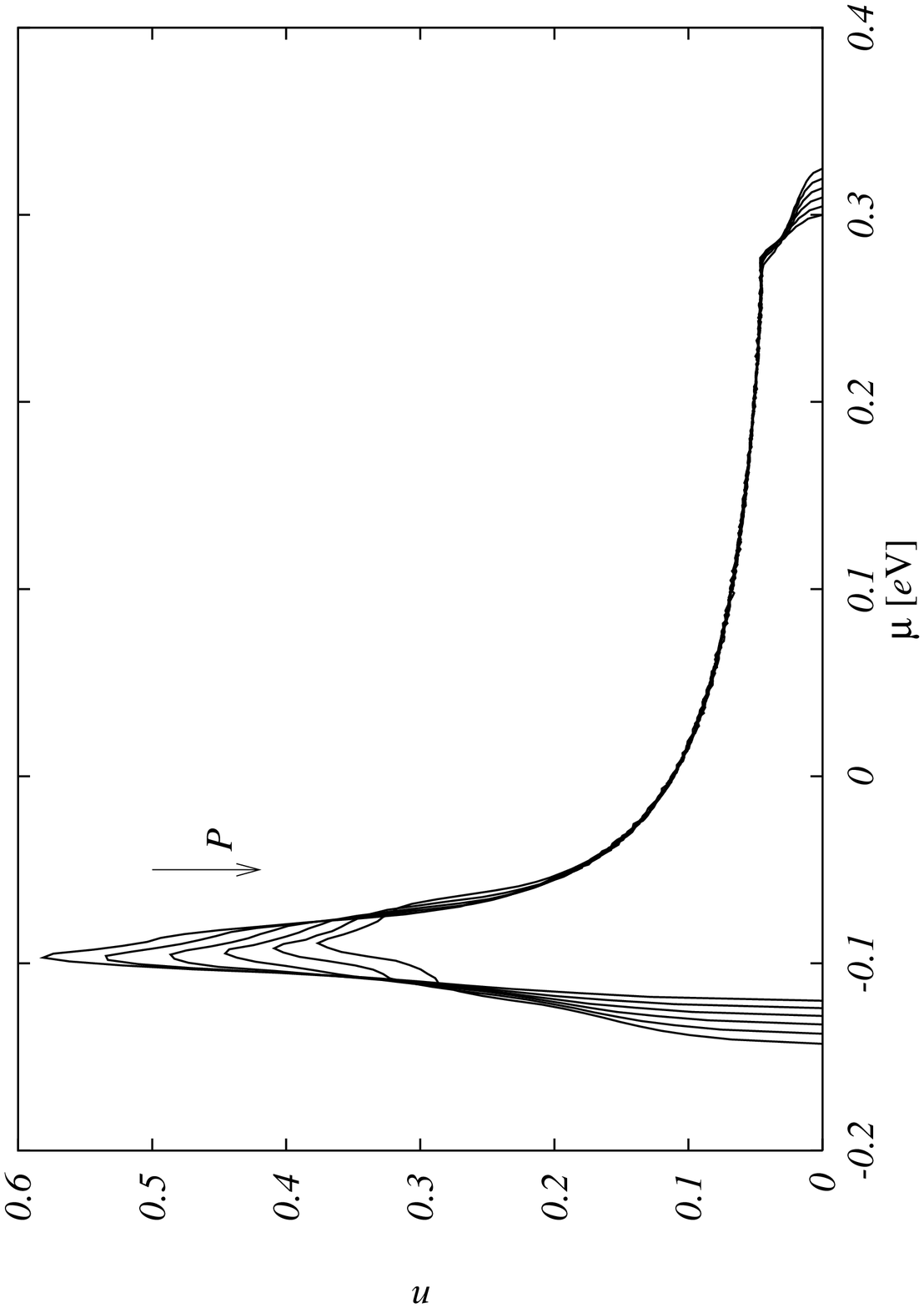}%
\end{center}
\caption{Density of states as a function of the Fermi level $\mu$,
$\mu_\bot \leq \mu \leq \mu_\top$. By increasing pressure $P=0.0\div 20.0$~GPa, the band
widens, while the DOS peak lowers.}
\label{fig:dens}
\efig

Let us now turn to the pressure dependence of the hole content.
Making the usual assumption that all the charge carriers are localized
within the Cu-O planes, one may identify the Hall resistance as
$R_{\rm H} = V/(ze\rho)$, where $e$ is the elementary charge and $z$
the number of Cu ions within a unit cell of volume $V$.
Equations~(\ref{eq:compressibilities}) then promptly yield
\begin{equation}
\kappa_\rho = \frac{1}{\rho} 
\left( \frac{\partial\rho}{\partial P} \right)_T =
\kappa_{\rm H} - \kappa_V ,
\end{equation}
where
\begin{equation}
\kappa_{\rm H} = -
\frac{1}{R_{\rm H}} \left( 
\frac{{\rm d} R_{\rm H}}{{\rm d} P} \right)_T .
\end{equation}
Therefore, at the lowest order in $P$,
\begin{equation}
\label{eq:rhoP}
\rho(P) = \rho(0) [1+\kappa_\rho P].
\end{equation}
However, one usually finds $\kappa_{\rm H} > \kappa_V$,
so that $\kappa_\rho >0$ for the majority of compounds.
For underdoped Bi2212, Huang~\cite{Huang}
et al. find $\kappa_{\rm H} =
+0.08$~GPa$^{-1}$, so that $\kappa_\rho =+0.0634$~GPa$^{-1}$.

In summary, Equation~(\ref{eq:rhoP})
fixes the hole content $\rho$ at the pressure $P$, given its value
$\rho(0)$ at normal pressure ($P=0$).
The knowledge of the lattice spacings, Eq.~(\ref{eq:aP}), and of the
band parameters as a function of $P$ yields a model $P$-dependent
band dispersion, giving rise to a flattened DOS (Fig.~\ref{fig:dens}). The
inversion of Eq.~(\ref{eq:rho}) then allows to evaluate the corresponding
chemical potential $\mu$, while Equation~(\ref{eq:discr})
eventually yields the critical temperature, $T_c$, as a function of the
coupling parameters, $U$, and $V$.
At this stage, we may leave the latter as free, and determine it in order 
to fit the known experimental dependence of $T_c$ on $P$.
Restricting to $d$-wave symmetry,
Equation~(\ref{eq:discr}) yields
\begin{equation}
\label{eq:V}
V = - \frac{1}{\langle D_{\bbox{k}}^2 \rangle_c}
\end{equation}
as a function of the critical temperature $T_c$.
Of course, nothing can be said about $U$, since its value does not affect
the $d$-wave solution for the gap.

\bfig
\begin{center}
\includifigura{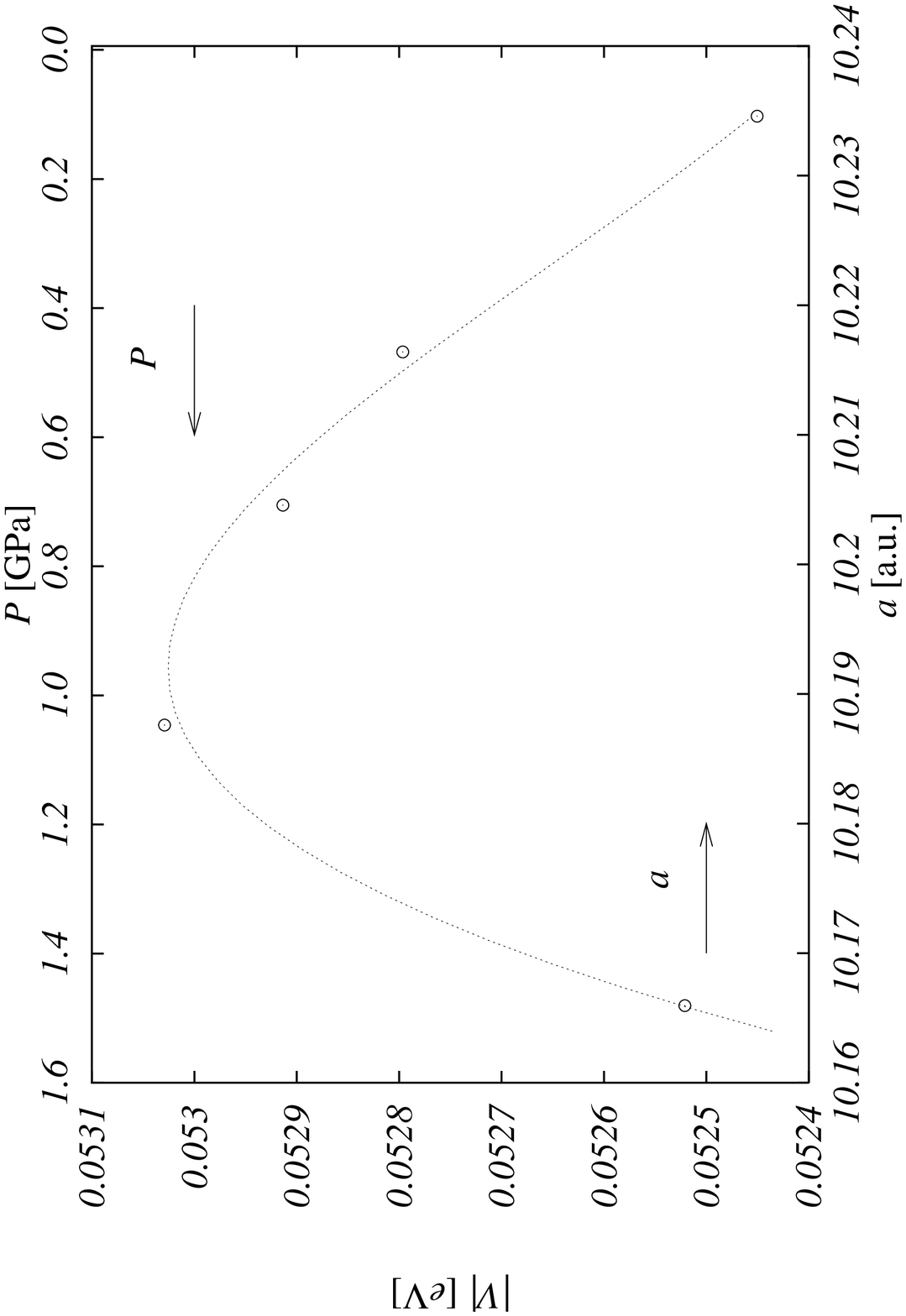}%
\end{center}
\caption{The absolute value of the inter-site coupling strength, $|V|$,
as a function of the in-plane lattice spacing $a=a_x \simeq a_y$ 
(bottom scale) and of pressure $P$ (top scale).
The circles make directly reference to the experimental
values $(P,\, T_c )$
given by Huang et al.,~\protect\cite{Huang} while the
dashed line has been evaluated by a standard best-fit procedure.
(cfr. also Fig.~\protect\ref{fig:dTc}).}
\label{fig:V}
\efig

Figure~\ref{fig:V} displays $V$, Eq.~(\ref{eq:V}), vs
$a_x$, Eq.~(\ref{eq:aP}), evaluated in correspondence
to the five experimental couples $(P,\, T_c )$ reported
for underdoped Bi2212 by Huang et al.~\cite{Huang}
The dashed line, which interpolates among these points, has been obtained by a
standard best-fit procedure for $P=0\div 1.6$~GPa.  Although the latter
curve yields an overall trend,
a nontrivial correlation is suggested between the interaction strength
and the structural properties of the lattice, such as its
in-plane spacing, which closely follows the dependence
of $T_c$ on $P$.~\cite{Huang} A similar conclusion has
been derived by Neumeier,~\cite{NeumeierC} using data
for the pressure dependence of $T_c$ in YBCO, 
and assuming a strong-coupling BCS in comparison with an
improved McMillan expression for $T_c$. Even if the latter
do not reliably account for the rich phenomenology of the
cuprates (viz., the high values of their $T_c$),
a satisfactory agreement was recognized with the observed trend
in ${\rm d}\ln T_c /{\rm d} P$. Both results seem to support a
nonspectator r\^ole of the lattice in the onset of an
electronic instability towards superconductivity.
Although the present analysis does not address the problem of the nature and origin
of the attractive interaction, Fig.~\ref{fig:V} suggests a non trivial behaviour of the
inter-site interaction $V$ as a function of the cell spacing.

We are eventually in the position to distinguish between the two contributions
to ${\rm d} T_c /{\rm d} P$ in Eq.~(\ref{eq:ansatz}). 
The solid line in Fig.~\ref{fig:dTc} 
is the best fit to the values of ${\rm d} T_c /{\rm d} P$ vs $T_c$
desumed from the experimental
work by  Huang et al~\cite{Huang}
on Bi2212 under pressures $P=0\div 1.6$~GPa.  Figure~\ref{fig:dTc}
also displays
\begin{equation}
\frac{\partial T_c}{\partial\rho} \frac{{\rm d}\rho}{{\rm d} P}
= \rho(0)\kappa_\rho \frac{\partial T_c}{\partial\rho} ,
\end{equation}
numerically evaluated on the basis of our results (Fig.~\ref{fig:Tc}),
assuming $\rho(0) \simeq0.2$ by comparison of 
Huang et al.'s results~\cite{Huang} with the known dependence
of $T_c$ on $\rho$ at normal pressure.~\cite{Allgeier}
The `intrinsic' contribution $\partial T_c /\partial P$ is
eventually resolved as the difference between the previous two.

A non-negligible `intrinsic' term $\partial T_c /\partial P$
is recognized, thus suggesting an effective
contribution of the lattice to the mechanism of high-$T_c$
superconductivity against the r\^ole of carriers. 
However, since $\partial T_c /\partial P >0$ in the pressure region
considered, the change of sign in the total ${\rm d} T_c / {\rm d} P$
observed in underdoped Bi2212 is mainly due to the non-monotonic dependence
of $T_c$ on $\rho$, via Eq.~(\ref{eq:ansatz}).

\bfig
\begin{center}
\includifigura{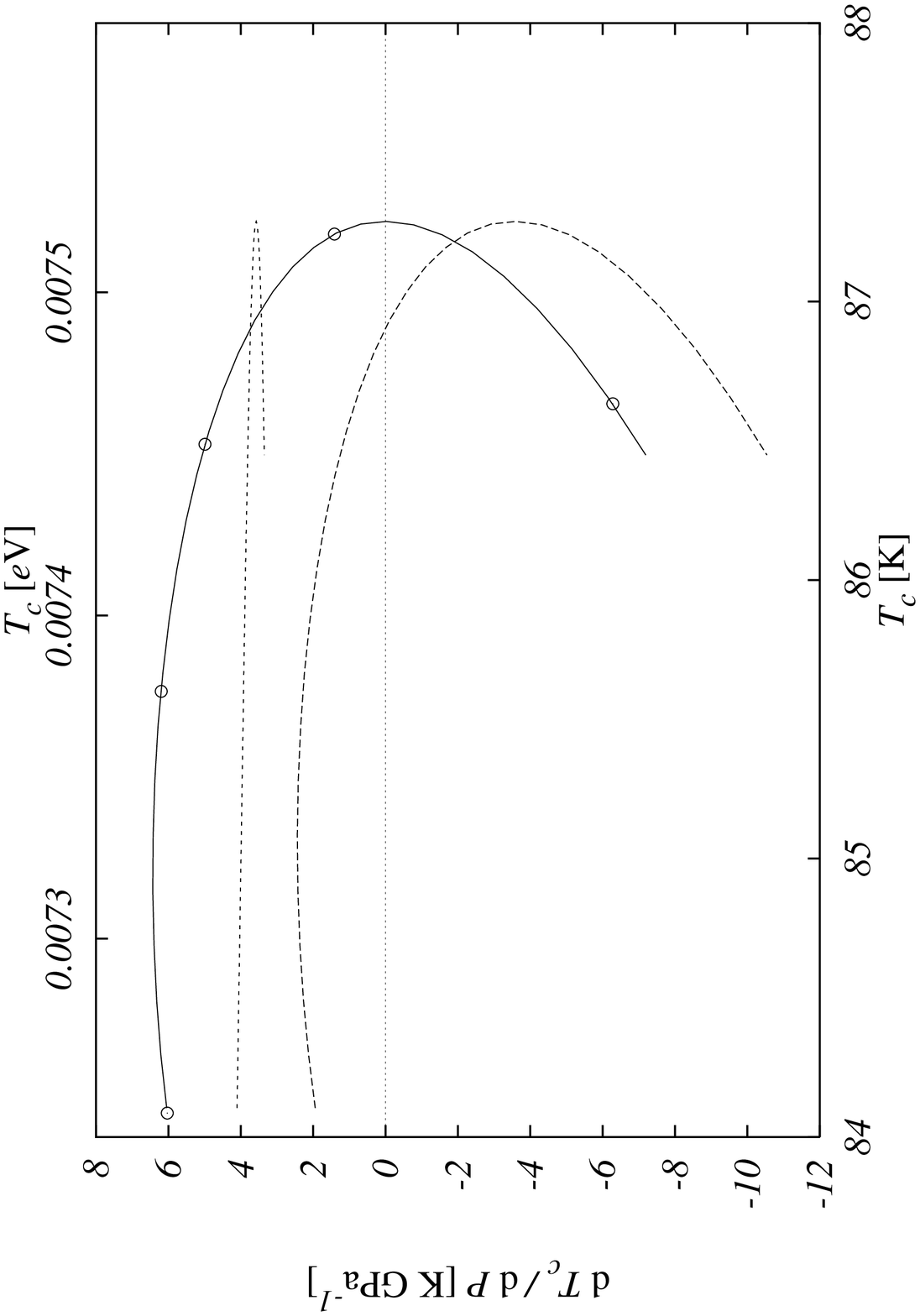}%
\end{center}
\caption{The solid line 
best fits Huang et al.'s experimental data~\protect\cite{Huang}
for ${\rm d} T_c /{\rm d} P$ vs $T_c$ in Bi2212
under pressures $P=0\div 1.6$~GPa (circles). Our numerical
estimates for the hole-induced (dashed line)
and `intrinsic' (dotted line) contributions to the
latter are also shown, according to Eq.~(\protect\ref{eq:ansatz}).}
\label{fig:dTc}
\efig

\section{Conclusions}

In summary, we have employed a BCS-type
mean-field approximation of the extended Hubbard 
model to describe a system of fermionic, hole-like carriers, subjected
to an intermediate-range interaction, in order to reproduce
the observed universal behaviour of $T_c$ vs the hole content $\rho$ in the high-$T_c$ superconductors.
A model tight-binding dispersion relation for the free carriers in
the lattice has been adopted, as suggested by ARPES measurements
and band structure calculations. A simplified, though accurate,
LCAO approximation has been employed in order to provide the band
parameters with a suitable dependence on the lattice steps,
and therefore on pressure, through the known compressibilities.
Due to the quasi-bidimensional lattice structure, 
an almost dispersionless character emerged for the carriers
in the direction orthogonal to the planes, and a quite large
van~Hove-like peak in their DOS. As predicted within an AFvH picture,
a pressure induced decrease in the DOS peak produces a shift
in the optimal doping level towards higher values.

The comparison with known experimental data for $T_c$ vs 
$\rho$,~\cite{Allgeier} and $T_c$ vs $P$~\cite{Huang}
in Bi2212 allowed us to interpret in a quantitative way the interplay
between $T_c$, $\rho$ and $P$. A strong
contribution to ${\rm d} T_c /{\rm d} P$, 
given by an increase of the hole
content through an applied pressure, evidenced by Hall resistance
pressure measurements, has been resolved from a
non-negligible `intrinsic' one,
mainly due to the lattice rearrangement
induced by pressure. On the basis of this result,
we can argue that the lattice structure is not completely
ininfluent on the mechanism of high-$T_c$ superconductivity. The
latter statement has been supported by a quantitative estimate
of a nontrivial correlation between the inter-site coupling
strength and the in-plane lattice spacings.

\acknowledgements
We gratefully acknowledge interesting discussions with Prof. J.S. Schilling
and Dr. A.-K. Klehe. We are furthermore indebted with
Dr. H. Takahashi and Dr. A. Nazarenko for providing us with
useful bibliography prior to its publication.
G.G.N.A. thanks C.N.R. and G.N.S.M. for financial support.

\appendix
\section*{Pressure dependence of the tight-binding parameters}
We here derive a convenient estimate of the dependence of the hopping parameters 
$t_x$, $t_y$, $t_{xy}$, $t_z$ which define the tight-binding model dispersion 
relation, Eq.~(\ref{eq:dispersion}), as a function of the lattice parameters,
$a_x$, $a_y$, $a_z$, and therefore of the pressure, through Eq.~(\ref{eq:aP}).

Parameters $t_x$, $t_y$ measure the probability for a carrier to hop from one
site to a nearest-neighbour site in the same Cu-O plane. In the tight-binding
approximation, the dependence of $t_x$, $t_y$ on the in-plane lattice spacings
$a_x$, $a_y$ may be approximated by the overlap integrals between
the $2p_x$, $2p_y$ hydrogenoid atomic orbitals, centered on the oxygen site,
and the $3d_{x^2 -y^2}$ one, centered on the copper site, distant
$a_x /2$, $a_y /2$ from the former, respectively.
Such overlap integrals have been analitically
evaluated employing elliptic coordinates, and they behave as $t_i
\approx \exp (-5a_i /24a_0  )$ for $a_i \gg a_0$, being $a_0$ the Bohr radius
of the hydrogen atom.

In an analogous way, we take $t_{xy}$ proportional to the overlap integral
between the $2p_x$ and the $2p_y$ hydrogenoid atomic orbitals, centered on
next-nearest-neighbours oxygen sites, respectively, distant $a_i \sqrt{2}$
apart. $t_z$ is assumed proportional to the overlap integral between the $3d_{3z^2
-r^2}$ hydrogenoid atomic orbital centered on a copper site, and the $2p_z$
one, centered on the corresponding apical oxygen. One finds
$t_{xy} \approx \exp (-a_{x,y} /2a_0 )$,
$t_z \approx \exp (-5 a_z /2 a_0 )$ for $a_i \gg a_0$.
The proportionality constants are chosen so that Eq.~(\ref{eq:dispersion})
correctly fits the observed band dispersion at normal pressure
(Fig.~\ref{fig:dispersion}).

\end{multicols}

\begin{references}
\bibitem{comment} The La-based compounds seem to be the main
exception to that rule, showing ${\rm d} \rho /{\rm d} P \approx 0$. 
Cfr. Ref.~[\negthinspace\negthinspace\onlinecite{Murayama}].
\end{references}
\end{document}